\documentstyle[11pt]{article}
\begin{document}
\begin{titlepage}
\title{\bf The Geometry of Self-dual 2-forms}
\date{ }
\author{ {\bf Ay\c{s}e H\"{u}meyra Bilge}\\{\small
Department of Mathematics, Institute for Basic Sciences}\\
{\small TUBITAK Marmara Research Center}\\
{\small Gebze-Kocaeli, Turkey}\\
{\footnotesize e.mail : bilge@yunus.mam.tubitak.gov.tr}\\
{\bf Tekin Dereli}\\{\small Department of Physics}\\
{\small Middle East Technical University}\\{\small Ankara, Turkey}
\\{\footnotesize e.mail : tdereli@rorqual.cc.metu.edu.tr}\\
{\bf \c{S}ahin Ko\c{c}ak}\\{\small Department of Mathematics}\\
{\small Anadolu University}\\{\small Eski\c{s}ehir, Turkey}\\
{\footnotesize e.mail : d170@vm.baum.anadolu.edu.tr} }
\maketitle
\begin{abstract}
{ We show that self-dual 2-forms in 2n dimensional spaces
determine a $n^2-n+1$
dimensional manifold ${\cal S}_{2n}$ and the dimension  of the maximal linear
subspaces of ${\cal S}_{2n}$ is equal  to the (Radon-Hurwitz) number of
linearly
independent vector fields on the sphere $S^{2n-1}$. We provide a direct
proof  that for $n$ odd ${\cal S}_{2n}$
has only one-dimensional linear
submanifolds. We exhibit $2^c-1$ dimensional subspaces in dimensions which
are multiples of $2^c$, for $c=1,2,3$. In particular, we demonstrate that
the  seven dimensional
linear subspaces of  ${\cal S}_{8}$ also include among many other interesting
classes of self-dual 2-forms,
the self-dual 2-forms of Corrigan, Devchand, Fairlie and Nuyts and  a
representation of ${\cal C}l_7$
 given by octonionic multiplication.
We discuss the relation of the linear subspaces with the representations of
Clifford algebras.}

 \end{abstract} \end{titlepage}
\noindent {\bf 1. Introduction}
\vskip 3mm

The self-dual Yang-Mills fields in four dimensions have
remarkable properties that have found several physical
applications. On the other hand
the notion of self-duality cannot be easily generalised to higher dimensions.
Here we  present a characterisation of (anti)self-dual
Yang-Mills fields
by an eigenvalue criterion. The main idea is given in our previous paper
${}^{[1]}$.
Here we study the geometry of the space of self dual 2-forms.

Let $M$ be a $2n$ dimensional differentiable
manifold, and $E$ be a vector bundle over
$M$ with standard fiber $R^n$ and  structure group $G$.
A Yang-Mills potential
can be represented by a $\cal G$ valued connection 1-form $A$ on $E$,
where $\cal G$ is a
linear representation of the Lie algebra of  the gauge group $G$. Then the
gauge fields are represented by the
curvature $F$ of the connection $A$
that is given locally by the $\cal G$ valued 2-form
$$F=dA-A\wedge A .$$
The Yang-Mills action is the $L_2$ norm of the
curvature 2-form $F$
$$\| F\|^2=\int_M {\rm tr} (F \wedge *F)$$
where $*$ denotes the Hodge dual defined relative to a positive definite
metric on $M$.
The Yang-Mills equations
$$d_EF=0,\quad ^*d_E^*F=0,$$
where $d_E$ is the bundle covariant derivative and
$-*d_E*$ is its formal adjoint, determine the critical points
of the action.
In $d=4$ dimensions
 $F$ is called self-dual or anti-self-dual provided
$$*F=\pm F.$$
Self-dual or anti-self-dual 2-forms are the global extrema of the Yang-Mills
action.  This can be seen as
follows:
The Yang-Mills action has a topological lower bound
$$\|F\|^2 \geq \int_M {\rm tr} (F\wedge F).$$
The term ${\rm tr} (F\wedge F)$ is related to the Chern classes
of the bundle. Actually if $E$ is a
complex 2-plane bundle with $ c_1(E)=0$, then
the topological bound is proportional to $ c_2(E)$ and
this lower bound is realised by a (anti)self-dual  connection.
Furthermore, $SU(2)$ bundles over a four manifold are classified by $\int
c_2(E)$, hence self-dual connections are minimal representatives of the
connections in each  equivalence class of $SU(2)$ bundles.
This is a generalisation of the fact that an $SU(2)$ bundle admits a flat
connection if and only if it is trivial.

In the literature essentially three notions of self-duality
in higher dimensions were being used.

i) A 2-form $F$ in dimension $2n$
is called self-dual if the Hodge dual of $F$ is proportional to $F^{n-1}$. ( Here  wedge  product of $F$'s should be understood.)
This notion is introduced by Trautman ${}^{[2]}$, and
Thcrakian ${}^{[3]}$ and used widely by others.
For details we refer to a review by Ivanova and Popov ${}^{[4]}$.

ii) Self-dual 2-forms $F$ in dimensions $2n = 4k$
are defined to be the ones such that $F^{k}$ is self-dual in the Hodge sense.
That is $^*F^k = \pm F^k$.
This is a nonlinear set of conditions and
the action which is minimised is $$\int_{M}tr(F^{k}\wedge ^*F^{k}).$$ This notion is adopted by Grossman, Kephardt and Stasheff (GKS) in their study of
instantons in eight dimensions ${}^{[5]}$.

iii) Both the criteria above  are non-linear.
Alternatively, (anti)self-dual 2-forms in $2n$ dimensions
can be defined as eigen-bivectors of a completely antisymmetric fourth rank tensor that is invariant under a subgroup of $SO(2n)$. The
set of such self-dual 2-forms
would span a linear space. This notion of self-duality is introduced
 by Corrigan, Devchand, Fairlie and Nuyts (CDFN )
who studied  the first-order  equations satisfied by  Yang-Mills  fields
in spaces of dimension
greater than four  and derived
$SO(7)$ self-duality equations in ${\bf R}^8$ ${}^{[6]}$.

It can be shown that the self-dual 2-forms defined by the above criteria
satisfy  Yang-Mills equations. However, the corresponding  Yang-Mills action need not be extremal.
In order to derive topological bounds in higher dimensions
we note that
the local curvature
2-form depends on a trivilization of the bundle, but its invariant
polynomials $\sigma_k$ defined by
$${\rm det}(I+tF)=\sum_{k=0}^n\sigma_k t^k$$
are invariant of the local trivialization. We recall that  $\sigma_k$ 's
are closed $2k$-forms,
hence they define  the
deRham cohomology classes in $H^{2k}$. Furthermore these
cohomology
classes depend only on the bundle.  They are  the Chern classes of the
bundle $E$ up to some multiplicative constants.
 It is also known that the $2k$-form $\sigma_k$ can be obtained as
linear combinations of $trF^k$ where $F^k$ means the product of the matrix $F$
with itself $k$ times, with the wedge multiplication of the entries.

In four dimensions the topological bound we wrote above is the only one that is available. In eight dimensions on the other hand
it is possible to introduce two independent topological bounds.
The topological lower bound on the action
$$\int_{M} tr(F^2 \wedge ^* F^2)  \ge k \int_M p_{2}(E)$$
corresponds to the choice of Thcrakian and GKS. The
self-duality (in the Hodge sense) of $F^2$ gives global minima of this
action involving the second Pontryagin number    $\int p_2(E)$.
In our previous paper ${}^{[1]}$ we introduced another
topological lower bound on the action
$$\int tr(F\wedge ^* F)^2\ge k' \int_{M}  p_1(E)^2.$$
This involves the square of the first Pontryagin number and has to be taken into account as the topology of the Yang-Mills bundle
on an eight manifold has to be characterised by both the first and the second  Pontryagin numbers.

The notion of self-duality introduced by us
 ${}^{[1]}$ encompasses all the criteria given above.
We recall here that a self-dual 2-form can be defined by an eigenvalue
criterion in the following way. ( We adopt a different
terminology, and use self-dual rather than strongly self-dual as
it is used in Ref.[1] )
Suppose $F$ is a real 2-form in 2n dimensions,
and let $\Omega$ be the corresponding $2n \times 2n$
skew-symmetric matrix with respect to some local orthonormal basis.
Then by a change of basis, $\Omega$ can be brought to the
block-diagonal form
$$\left ( \begin{array}{ccccccc}
0&\lambda_{1}& & & & & \\
-\lambda_{1}&0& & & & & \\ & &. & & & & \\ & & &. & & & \\ & & & &.& &  \\
& & & &  & 0&\lambda_{n}\\ & & & &  &- \lambda_{n}&0
\end{array} \right )
$$
where $\lambda_{1},...,\lambda_{n}$ are the eigenvalues of $\Omega$.
The 2-form $F$ is called   self-dual or anti-self-dual provided
the absolute values of the eigenvalues are all equal , that is
$$|\lambda_{1}|=|\lambda_{2}|= \dots =|\lambda_{n}|.$$
To distinguish between the two cases, orientation must be taken into account.
We define $F$ to be self-dual, if $\Omega$ can be brought with respect
to an orientation-preserving basis change to the above block-diagonal
form such that $ \lambda_{1}=\lambda_{2}= \dots = \lambda_{n}$.
Similarly, we define $F$ to be anti-self-dual, if $\Omega$
can be brought to the same form  by an orientation-reversing basis
change.
It is not difficult to check that in dimension D=4, the above definition
coincides with the usual definition of self-duality in the Hodge sense.

We have already shown that the definition of self-duality by the equality of the eigenvalues implies the criteria (i) and (ii), and  the CDFN 2-forms
in eight dimensions are self-dual in the above sense.

Let ${\cal S}_{2n}$ be the set of self-dual 2-forms in $2n$ dimensions.
In Section 2 we give the manifold structure of ${\cal S}_{2n}$. In Section 3, we show that the dimension of maximal linear spaces of ${\cal S}_{2n}$ is equal to the number of linearly independent vector fields on $S^{2n-1}$.
We give a direct proof that in eight dimensions
%%
%first derived by Corrigan, Devchand, Fairlie, Nuyts (CDFN) ${}^{[2]}$
%by other means, are  self-dual in the above sense.
%ii) Each self-dual 2-form $F$, satisfying
%$ \ast(F \wedge F) = F \wedge F$
%${}^{[3]}$ is self-dual in the above sense.
%iii)
%
 starting from the self-duality condition on eigenvalues
we obtain the CDFN self-dual 2-form.
We also explain the construction of new  families of
self-dual 2-forms in ${\cal S}_8$
in terms of Clifford representations using octonionic
multiplication.
 \vskip 0.5cm
\noindent
{\bf 2. The Geometry of Self-dual 2-forms.}
\vskip .2cm

In this section we describe the geometrical structure of  self-dual
2-forms in arbitrary even dimensions. $I$ denotes
an identity
matrix of appropriate dimension.
 \vskip .1cm
 \noindent
{\bf Definition 1.} Let ${\bf A}_{2n}$ be the set of antisymmetric matrices in
$2n$ dimensions. Then ${\cal S}_{2n}=\{ A\in {\bf A}_{2n}
\mid A^2+\lambda^2I=0,\lambda\in {\bf R}, \lambda \ne 0\}$.

Note that if $A\in {\cal S}_{2n}$, and $A^2=0$, then $A=0$, and if
$A\in {\cal S}_{2n}$,
then $\lambda A\in{\cal S}_{2n}$ for $\lambda \ne 0$.

\vskip .2cm

\noindent
\proclaim Proposition 2. The set ${\cal S}_{2n}$ is diffeomorphic to
$\big(O(2n)\cap{\bf A}_{2n}\big)\times {\bf R}^+$.

\noindent
{\it Proof.} Let $A\in{\cal S}_{2n}$ with $A^2+\lambda^2I=0$. Note that
$\lambda^2=-\frac{1}{2n}{\it tr}A^2$.  Define $\kappa=\big[-\frac{1}{2n}{\it
tr}A^2\big]^{1/2}$, $\tilde{A}=\frac{1}{\kappa}A$. Then, $\tilde{A}^2+I=0$,
hence $\tilde{A}\tilde{A}^\dagger=I$. Consider the map
$\varphi:{\cal S}_{2n}\to
\big(O(2n)\cap {\bf A}_{2n}\big)\times {\bf R}^+$ defined by $\varphi(A)=(\tilde
{A},\kappa)$. The map $\varphi $ is 1-1, onto and differentiable. Its inverse
is given by $(B,\alpha)\to\alpha B$ is also differentiable, hence $\varphi$ is
a diffeomorphism.\quad e.o.p.

\vskip .2cm
\noindent
{\bf Remark 3.}
$ O(2n) \cap {\bf A}_{2n}$ is a fibre bundle over the sphere
$S^{2n-2}$ with fibre $O(2n-2) \cap {\bf A}_{2n-2}$.
(See Steenrod, Ref.{[7]})
\vskip .1cm
For our purposes the following description of ${\cal S}_{2n}$
is more useful.

\proclaim Proposition 4. ${\cal S}_{2n}$ is diffeomorphic to the homogeneous
manifold $\big(O(2n)\times {\bf R}^+\big)/U(n) \times \{ 1 \}$,
and {\it dim}${\cal S}_{2n} =n^2-n+1$.

 \vskip .2cm
\noindent
{\it Proof.} Let $G$ be the product group $O(2n)\times {\bf R}^+$,
where ${\bf R}^+$ is
considered as a multiplicative group. $G$ acts on ${\cal S}_{2n}$
by $(P,\alpha)\dot
A=\alpha(P^t A P)$, where $P\in O(2n)$, $\alpha\in {\bf R}^+$,
$A\in{\cal S}_{2n}$, and $t$
indicates the transpose. Since all matrices in ${\cal S}_{2n}$
are conjugate to each
other up to a multiplicative constant, this action is transitive, and actually
any $A\in {\cal S}_{2n}$ can be written as $A=\lambda P^t JP$,
where $\lambda=
\big[-\frac{1}{2n}{\it tr} A^2\big]^{1/2}$, with $P\in O(2n) $ and
$J=\pmatrix{0&I\cr -I&0\cr}$. It can be seen that the isotropy subgroup of $G$
at $J$ is $U(n)$ ${}^{[8]}$ and $G/U(n)$ is diffeomorphic to ${\cal S}_{2n}$
( Ref.{[9]} p.132, Thm.3.62 )  Then {\it dim}${\cal S}_{2n}=dim\big(O(2n)\times
{\bf R}^+/U(n)\big)$ can be easily computed as
{\it dim}${\cal S}=dim O(2n)+1-dimU(n)=(2n^2-n+1)-n^2=n^2-n+1$.
\quad e.o.p.

In particular, in eight dimensions, ${\cal S}_8$ is a 13 dimensional
manifold.

As $O(2n)$ has two connected components ( $SO(2n)$ and $O(2n) \setminus
SO(2n)$), $U(n)$ is connected and $U(n) \subset SO(2n)$,
${\cal S}_{2n}$ has two connected components. One of them
(that contains J) consists of the self-dual forms and the other of the
anti-self-dual forms.

\vskip .5cm
\noindent
{\bf 3. Maximal linear submanifolds of ${\cal S}_{2n}$ }
\vskip .2cm

In this section we will show that the  dimension of maximal linear subspaces of
${\cal S}_{2n}$ is  equal to  the number of linearly independent vector fields
on $S^{2n-1}$. The maximal number  of pointwise linearly independent vector
fields on the sphere $S^N$ is given by the Radon-Hurwitz number $k$. If
$N+1=2n=(2a+1)2^{4d+c}$ with $c=0,1,2,3$, then $k=8d+2^c-1$
( See e.g. Ref. [10],p.45, Thm.7.2). This construction gives
three vector fields on $S^3$, seven on $S^7$, three on $S^{11}$,
eight on $S^{15}$ and so on. In particular
there is only one vector field on the spheres $S^{2n-1}$ for odd n.

Let ${\cal L}_{2n}^\alpha$ be a maximal linear subspace  of ${\cal S}_{2n}$,
where
$\alpha$ is a real parameter. Since the elements of ${\cal L}_{2n}^\alpha $
are
skew-symmetric and non-degenerate, the dimension of ${\cal L}_{2n}^\alpha$ is
less than or equal to $2n-1$.
For example in dimension eight ${\cal S}_8$ is $13$ dimensional, and we will
show that the maximal linear subspaces are $7$ dimensional, hence they form six dimensional families.

\proclaim Proposition 5. The dimension of the maximal linear subspaces of
${\cal S}_{2n}$ is  equal to the number of linearly independent
vector fields on $S^{2n-1}$.

\vskip .2cm
\noindent
{\it Proof.} We will show  that the bases of linear subspaces of ${\cal
S}_{2n}$ give rise to linearly  independent vector fields on $S^{2n-1}$.
Let $\{h_i\}$, $i=1,\dots k$ be  an orthogonal basis for ${\cal
L}_{2n}^\alpha$. That is the $h_i$'s are  linearly independent matrices
satisfying  ${\rm tr}(h_i^th_j)=\delta_{ij}$.  Suppose $\xi_1,\dots,\xi_{2n}$
are coordinates on $R^{2n}$ and let $R=(\xi_1,\dots,\xi_{2n})$ be the radial
vector in $R^{2n}$. Define the vector fields $X_i=h_iR$. Then $X_i$'s are the
tangent vector fields to $S^{2n-1}$, since $\langle X_i,R\rangle=R^th_iR=0$, by
the skew-symmetry of $h_i$'s. The linear independence of $h_i$'s implies the
linear independence of the $X_i$'s and thus proves our claim:
                                                              $\sum _{i=1}^k
\lambda_iX_i=\sum _{i=1}^k\lambda_i h_iR=0$ implies
$\lambda_1=\dots=\lambda_k=0$ because $h_i$'s are linearly independent.

This shows that the dimension of a  maximal linear subspace of ${\cal S}_{2n}$
is less than the Radon-Hurwitz number. Conversely, the Radon-Hurwitz number $k$
(associated to $2n$) is equal to the maximal dimension of the Clifford algebra
acting on $R^{2n}$ [10]. If we take such a representation of ${\cal C}l_k$ on
$R^{2n}$, the images of a generator set $\{v_1,\dots, v_k\}$ (with $v_i^2=-1$,
$ v_iv_j+v_jv_i=0$ for $i\ne j$) are given by skew-symmetric matrices with
respect to an appropriate basis of $R^{2n}$. These images generate linearly a
$k$-dimensional subspace of ${\cal S}_{2n}$. This shows that the dimension of a
maximal linear subspace of ${\cal S}_{2n}$ is equal to the Radon-Hurwitz
number. \quad e.o.p.

\vskip .2cm
This property shows that there is an intimate relationship between
generalised self-duality and Clifford algebras. We will give a systematic
exposition of this relationship in a subsequent publication.

We remark  that  $X_i$'s  form an orthogonal  frame. As
$h_i$'s and
$(h_i+h_j)$'s both belong to  ${\cal S}_{2n}$, $h_i^2=-k_i^2I$, and
$h_ih_j+h_jh_i=k_{ij}I$ for some  constants $k_i$ and $k_{ij}$. Then since
$\langle h_i,h_j\rangle={\rm tr}(h_i^t,hj)=0$ and trace is symmetric, it
follows that $h_ih_j+h_jh_i=0$. Then
\begin{eqnarray}
2\langle X_i,X_j\rangle&=&\langle X_i,X_j\rangle+\langle X_j,X_i\rangle\cr
  &=&R^t(h_i^th_j+h_j^th_i)R\cr
  &=&-R^t(h_ih_j+h_jh_i)R\cr
  &=&0. \nonumber
\end{eqnarray}

We now directly prove that for odd n there are no linear subspaces
other than the one dimensional ones.
\proclaim Proposition 6. Let ${\cal M}=\{A\in {\cal S}\mid (A+J_o)\in {\cal
S}\}$. Then ${\cal M}= \{ k J | k \epsilon {\bf R} \}$ for odd $n$.

\vskip .2cm
\noindent
{\it Proof.}
Let $A=\pmatrix{A_{11}&A_{12}\cr
-A_{12}^t&A_{22}\cr}$, where $A_{11}+A_{11}^t=0$, $A_{22}+A_{22}^t=0$.
As before if $(A+J_o)\in {\cal S}$ then $AJ_o+J_oA$ is
proportional to the identity.
This gives
$A_{11}+A_{22}=0$ and the symmetric part of $A_{12}$ is proportional to
identity. Therefore $A=kJ_o+\pmatrix{A_{11}&A_{12o}\cr A_{12o}&-A_{11}\cr}$,
where $A_{12o}$ denotes the antisymmetric part of $A_{12}$ and $k$ is a
constant. Then the requirement that $A\in {\cal S}$ gives
$$[A_{11},A_{12o}]=0,\quad A_{11}^2+A_{12o}^2+kI=0,\quad k\in R.$$
As $A_{11}$ and $A_{12o}$ commute, they can be simultaneously diagonalisable,
hence for odd $n$ they can be brought to the form
$$A_{11}=diag(\lambda_1\epsilon,\dots,\lambda_{(n-1)/2}\epsilon,0)$$
$$A_{12o}=diag(\mu_1\epsilon,\dots,\mu_{(n-1)/2}\epsilon,0)$$
where $\epsilon=\pmatrix{0&1\cr-1&0\cr}$, and $0$ denotes a $1\times
1$ block, up to the permutation of blocks. If the blocks occur as shown,
clearly $A_{11}^2+A_{12o}^2$ cannot be proportional to identity.
It can also be
seen that except for $\lambda_i = \mu_i =0$
the same result holds for any permutation of the blocks. \quad e.o.p.
\vskip 5mm

\noindent
{\bf 4. An explicit construction of linear submanifolds of ${\cal S}_{8}$ }
\vskip .2cm

The defining equations of the set ${\cal S}_8$ are homogeneous quadratic
polynomial
equations for the components of the curvature 2-form and they correspond
to differential equations which are quadratic in the first derivative for the
connection. Thus the study of their solutions, hence the study of the moduli
space of  self-dual
connections is rather difficult. On the other hand the self-dual 2-forms lying
in a linear subspace of ${\cal S}_{2n}$ will correspond to linear gauge field
equations.
The study of the structure of the linear submanifolds of ${\cal S}_{2n}$ in
general is not attempted here, but at least for ${\cal S}_8$ we know that these
linear submanifolds form a 6-parameter family, and  there is no
a priori reason to single out one of them.

In Ref.{[1]} we
have shown that the 2-forms satisfying a set of 21 equations proposed by
Corrigan et al belong to  ${\cal S}_8$. We shall first give a natural
way of arriving at them, but it will depend on a reference form.
Changing the reference form one obtains translates of this 7-dimensional
plane, which in some cases look more pregnant than the original one.
Then we shall give a general procedure to construct self-dual 2-forms in $4n$
dimensions using  self-dual/anti self-dual forms and certain symmetric
matrices
in $2n$ dimensions. The matrices corresponding to these building blocks are
actually the representations of Clifford algebras in the skew-symmetric
matrices and dual Clifford algebras in symmetric matrices in half dimensions.
The CDFN plane, and the representation
   of ${\cal C}l_7$ using octonionic multiplication
 will arise naturally from these constructions.

Note that we excluded the zero matrix from ${\cal S}_{2n}$
in our definition in order
to obtain its manifold structure. We denote $\overline{\cal S}_{2n}
={\cal S}_{2n} \cup
\{0\}$. By linearity of the action of $O(2n) $ on ${\cal S}_{2n}$
we obtain the following

\proclaim Lemma 7. Let ${\cal L}$ be a linear submanifold of $\overline{{\cal
S}}_{2n}$. Then
${\cal L}_P=P^t {\cal L} P$, $P\in O(2n)$ is also a linear submanifold of
$\overline{{\cal S}}_{2n}$.

Let $J_o=diag(\epsilon,\epsilon,\epsilon,\epsilon)$, where
$\epsilon=\pmatrix{0&1\cr-1&0\cr}$. Note that any $A\in {\cal S}_8$
is conjugate
to $J_o$, hence any linear subset of $\overline{{\cal S}}_8$ can be realized as
the translate of a linear submanifold containing $J_o$ under conjugation.
Thus without loss of generality we can concentrate on linear subsets
containing $J_o$.
In the following, by abuse of notation we will not distinguish between ${\cal
S}_{2n}$ and  its closure.

 \proclaim Proposition 8. If $A\in {\cal S}_8$ and $(A+J_o)\in
{\cal S}_8$, where
$J_o=diag(\epsilon,\epsilon,\epsilon,\epsilon)$, with
$\epsilon=\pmatrix{0&1\cr-1&0\cr}$, then
$$A=\pmatrix{k\epsilon &r_1S(\alpha) &r_2S(\beta)   &r_3S(\gamma)\cr
        -r_1S(\alpha)   &k\epsilon    &r_3S(\gamma') &-r_2S(\beta')\cr
        -r_2S(\beta)   &-r_3S(\gamma')&k\epsilon    &r_1S(\alpha')\cr
        -r_3S(\gamma)  &r_2S(\beta')  &-r_1S(\alpha')&k\epsilon\cr}$$
where $k\in R$, $r_1$, $r_2$, $r_3$ are in $R^+$, and
$S(\theta)=\pmatrix{\cos \theta & \sin \theta\cr \sin \theta &-\cos
\theta\cr}$, and $\alpha$, $\alpha'$, $\beta$, $\beta'$, $\gamma$, $\gamma'$
satisfy
 $$\alpha+\alpha'=\beta+\beta'=\gamma+\gamma'$$.

\vskip .2cm
\noindent
{\it Proof.} If $A$ and $A+J_o$ are both in ${\cal S}_8$, then the matrix
$AJ_o+J_oA$ is proportional to identity. This gives a set of linear equations
whose solutions can be obtained without difficulty
to yield
$$A=\pmatrix{
      0& a_{12}& a_{13}& a_{14}& a_{15}& a_{16}& a_{17}& a_{18}\cr
-a_{12}&      0& a_{14}&-a_{13}& a_{16}&-a_{15}& a_{18}&-a_{17}\cr
-a_{13}&-a_{14}&      0& a_{12}& a_{35}& a_{36}& a_{37}& a_{38}\cr
-a_{14}& a_{13}&-a_{12}&      0& a_{36}&-a_{35}& a_{38}&-a_{37}\cr
-a_{15}&-a_{16}&-a_{35}&-a_{36}&      0& a_{12}& a_{57}& a_{58}\cr
-a_{16}& a_{15}&-a_{36}& a_{35}&-a_{12}&      0& a_{58}&-a_{57}\cr
-a_{17}&-a_{18}&-a_{37}&-a_{38}&-a_{57}&-a_{58}&      0& a_{12}\cr
-a_{18}& a_{17}&-a_{38}& a_{37}&-a_{58}& a_{57}&-a_{12}&      0\cr}$$
 Then the requirement that the diagonal entries in $A^2$ be
equal to each other give the following equations after some
algebraic manipulations:
$$a_{13}^2+a_{14}^2=a_{57}^2+a_{58}^2$$
$$a_{15}^2+a_{16}^2=a_{37}^2+a_{38}^2$$
$$a_{17}^2+a_{18}^2=a_{35}^2+a_{36}^2$$
Thus we can parametrise $A$ by
$$a_{13}=r_1\cos \alpha,\quad a_{15}=r_2\cos\beta\quad a_{17}=r_3\cos \gamma$$
$$a_{14}=r_1\sin \alpha,\quad a_{16}=r_2\sin\beta\quad a_{18}=r_3\sin \gamma$$
$$a_{57}=r_1\cos \alpha',\quad a_{37}=r_2\cos\beta'\quad
  a_{35}=r_3\cos \gamma'$$
$$a_{58}=r_1\sin \alpha',\quad a_{38}=r_2\sin\beta'\quad
  a_{36}=r_3\sin \gamma'$$

Finally the requirement that the off diagonal terms in $A^2$ be equal to zero
gives quadratic equations, which can be rearranged and using trigonometric
identities they give
$\alpha+\alpha'=\beta+\beta'=\gamma+\gamma'$. \quad e.o.p.

Thus the set of matrices $A\in {\cal S}_8$ such that
$(A+J_o)\in {\cal S}_8$ constitutes
an eight parameter family and  the
equations of CDFN  correspond to the case $\alpha'+\alpha = \beta'+\beta =
\gamma'+\gamma=0 $.
Thus we have an invariant description of these equations, that we repeat here
for convenience.
\begin{eqnarray}
&F_{12}-F_{34}=0\quad F_{12}-F_{56}=0\quad F_{12}-F_{78}=0\cr
&F_{13}+F_{24}=0\quad F_{13}-F_{57}=0\quad F_{13}+F_{68}=0\cr
&F_{14}-F_{23}=0\quad F_{14}+F_{67}=0\quad F_{14}+F_{58}=0\cr
&F_{15}+F_{26}=0\quad F_{15}+F_{37}=0\quad F_{15}-F_{48}=0\cr
&F_{16}-F_{25}=0\quad F_{16}-F_{38}=0\quad F_{16}-F_{47}=0\cr
&F_{17}+F_{28}=0\quad F_{17}-F_{35}=0\quad F_{17}+F_{46}=0\cr
&F_{18}-F_{27}=0\quad F_{18}+F_{36}=0\quad F_{18}+F_{45}=0\cr \nonumber
\end{eqnarray}
The (skew-symmetric) matrix of such a 2-form is
$$ \left ( \begin{array}{cccccccc}
0& F_{12}&F_{13}&F_{14}&F_{15}&F_{16}&F_{17}&F_{18}\\
 &  0&F_{14}& -F_{13}&F_{16 }&-F_{15}&F_{18}&-F_{17}\\
 & &  0 &F_{12}& F_{17}&-F_{18}&-F_{15}&F_{16}\\
 & & & 0& -F_{18}&-F_{17}&F_{16}&F_{15}\\
 & & & & 0&F_{12}&F_{13}&-F_{14}\\
 & & & & & 0&-F_{14}&-F_{13}\\
 & & & & & & 0 & F_{12}\\
 & & & & & & &  0 \end{array} \right ) $$
We will  refer to the plane consisting of these forms as the
CDFN-plane. Let us now consider as the reference form
$J=\pmatrix{0&I \cr -I&0 \cr }$ instead of $J_o$.
$J$ can be obtained from $J_o$ by conjugation $ J = P^t J_o P$ with
$$P= \left ( \begin{array}{cccccccc}
1&0&0&0&0&0&0&0\\
0&0&0&0&1&0&0&0\\
0&1&0&0&0&0&0&0\\
0&0&0&0&0&1&0&0\\
0&0&1&0&0&0&0&0\\
0&0&0&0&0&0&1&0\\
0&0&0&1&0&0&0&0\\
0&0&0&0&0&0&0&1 \end{array} \right ) $$
Then the conjugation of the CDFN-plane by $P$ is given by the following
( D=8 self-dual) 2-form
$$ F_{12} J + \pmatrix{ \Omega'& \Omega''\cr {\Omega''} & -\Omega'\cr}$$
where $\Omega'$ is a D=4 self-dual 2-form and $\Omega''$ is a D=4
anti-self-dual 2-form.
We found it remarkable that a similar construction was given
a long time ago by Witten${ }^{[11]}$.

At the end of this section, we shall obtain this
plane from a general rule for
the construction of orthonormal bases for linear subspaces and also show
that it corresponds to the representation of  $Cl_7$ ising octonionic
multiplication.

We shall now discuss a general procedure for constructing  linear subspaces of
self-dual forms.
Note that ${\cal S}_{2n}$ is the set skew-symmetric matrices
in $O(2n)\times R$. We define ${\cal P}_{2n}$ to be the set of symmetric
matrices in $O(2n)\times R$. Recall that an orthonormal basis for a
$k$-dimensional linear subspaces of ${\cal S}_{2n}$ corresponds to
the representation of $Cl_k$ in the skew-symmetric matrices. Similarly an
orthonormal  basis
for a $k$-dimensional linear subspace of ${\cal P}_{2n}$ corresponds to a
representation of the dual  Clifford algebra  $Cl'_k$ in the symmetric
matrices. These
bases will be the building blocks for self-dual forms in the double dimension.

We have already shown that in dimensions $2n=2(2a+1)$ the maximal  linear
subspaces of ${\cal S}_{2n}$ were one dimensional. Similarly, in
dimensions
$2n=4(2a+1)$, the dimension of maximal linear subspaces of ${\cal S}_{2n}$ are
three dimensional. It can be seen that the matrices
$$J_0=\pmatrix{ 0& 0& I& 0\cr
                0& 0& 0& I\cr
               -I& 0& 0& I\cr
                0&-I& 0& 0\cr},\quad
  J_1=\pmatrix{ 0& I& 0& 0\cr
               -I& 0& 0& 0\cr
                0& 0& 0&-I\cr
                0& 0& I& 0\cr},\quad
  J_2=\pmatrix{ 0& 0& 0& I\cr
                0& 0&-I& 0\cr
                0& I& 0& 0\cr
               -I& 0& 0& 0\cr},\quad$$
where $I$ is the identity matrix, form an orthonormal basis
for three dimensional linear subspaces of ${\cal S}_{4(2a+1)}$.

From now on we consider the self-dual 2-forms in $8n$ dimensions.
The matrix of a self-dual form can be written in the form
$$F=\pmatrix{A_a&B_a+B_s\cr
             B_a-B_s&D_a\cr},$$
where the matrices $A_a$, $B_a$, $D_a$'s are anti-symmetrical and  $B_s$ is
symmetrical. The requirement that $F^2$ be proportional to
the identity matrix gives the following equations:
$$           A_a^2= D_a^2,\quad\quad A_a^2+B_a^2-B_s^2=kI,\quad\quad
               [B_a,B_s]=0,$$
$$           A_aB_a +B_aD_a=0,\quad \quad B_aA_a+D_aB_a=0,$$
$$           A_aB_s +B_sD_a=0,\quad \quad B_sA_a+D_aB_s=0.$$
Now if we furthermore require that $F$ be build up from the linear
subspaces of ${\cal S}_{4n}$ and ${\cal P}_{4n}$, then we see that $A_a$,
$D_a$, $B_a$, $B_s$ have to be nondegenerate.

We shall give now an explicit construction of various linear subspaces of
${\cal S}_8$. Let ${\cal A}^-$ and ${\cal A}^+$ be orthonormal bases for linear
subspaces of ${\cal S}_{2n}$ and ${\cal P}_{2n}$, respectively.

In two dimensions we have the following structure.
$${\cal A}^-=\left\{\pmatrix{0&1\cr-1&0\cr}\right\},\quad
  {\cal A}_{(1)}^+=\left\{\pmatrix{1&0\cr0&1\cr}\right\},\quad
  {\cal
A}_{(2)}^+=\left\{\pmatrix{1&0\cr0&-1\cr},\pmatrix{0&1\cr1&0\cr}\right\}.$$

  From the commutation relations it can be seen that the orthonormal
bases for linear subspaces of self-dual 2-forms in four dimensions are
determined by the choice of $B_s$. The choice $B_s\in{\cal A}_{(1)}^+$ leads
to the usual anti self-dual 2-forms, while the choice $B_s\in{\cal A}_{(2)}^+$
leads to the self-dual 2-forms.  Hence in four dimensions we obtain two
different sets of orthonormal bases for linear subspaces of ${\cal S}_4$.
By similar considerations, we obtain seven different bases for linear
subspaces of ${\cal P}_4$. The elements of these bases are listed below:

$$
a_1=\pmatrix{ 0& 1 & 0 & 0\cr
             -1& 0 & 0 & 0\cr
              0& 0 & 0 & 1\cr
              0& 0 &-1 & 0\cr},\quad
a_2=\pmatrix{ 0& 0 & 1 & 0\cr
              0& 0 & 0 &-1\cr
             -1& 0 & 0 & 0\cr
              0& 1 & 0 & 0\cr},\quad
a_3=\pmatrix{ 0& 0 & 0 & 1\cr
              0& 0 & 1 & 0\cr
              0&-1 & 0 & 0\cr
             -1& 0 & 0 & 0\cr},\quad$$
$$b_1=\pmatrix{ 0& 1 & 0 & 0\cr
             -1& 0 & 0 & 0\cr
              0& 0 & 0 &-1\cr
              0& 0 & 1 & 0\cr},\quad
b_2=\pmatrix{ 0& 0 & 1 & 0\cr
              0& 0 & 0 & 1\cr
             -1& 0 & 0 & 0\cr
              0&-1 & 0 & 0\cr},\quad
b_3=\pmatrix{ 0& 0 & 0 & 1\cr
              0& 0 &-1 & 0\cr
              0& 1 & 0 & 0\cr
             -1& 0 & 0 & 0\cr},\quad$$
$$c_1=\pmatrix{ 0& 0 & 0 & 1\cr
              0& 0 &-1 & 0\cr
              0&-1 & 0 & 0\cr
              1& 0 & 0 & 0\cr},\quad
c_2=\pmatrix{ 0& 0 & 1 & 0\cr
              0& 0 & 0 & 1\cr
              1& 0 & 0 & 0\cr
              0& 1 & 0 & 0\cr},\quad
p_1=\pmatrix{ 0& 1 & 0 & 0\cr
              1& 0 & 0 & 0\cr
              0& 0 & 0 & 1\cr
              0& 0 & 1 & 0\cr},\quad$$
$$p_2=\pmatrix{ 1& 0 & 0 & 0\cr
              0&-1 & 0 & 0\cr
              0& 0 & 1 & 0\cr
              0& 0 & 0 &-1\cr},\quad
d_1=\pmatrix{ 0& 0 & 0 & 1\cr
              0& 0 & 1 & 0\cr
              0& 1 & 0 & 0\cr
              1& 0 & 0 & 0\cr},\quad
d_2=\pmatrix{ 0& 0 & 1 & 0\cr
              0& 0 & 0 &-1\cr
              1& 0 & 0 & 0\cr
              0&-1 & 0 & 0\cr},\quad$$
$$q_1=\pmatrix{ 0& 1 & 0 & 0\cr
              1& 0 & 0 & 0\cr
              0& 0 & 0 &-1\cr
              0& 0 &-1 & 0\cr},\quad
q_2=\pmatrix{ 1& 0 & 0 & 0\cr
              0&-1 & 0 & 0\cr
              0& 0 &-1 & 0\cr
              0& 0 & 0 & 1\cr},\quad
e_1=\pmatrix{ 1& 0 & 0 & 0\cr
              0& 1 & 0 & 0\cr
              0& 0 &-1 & 0\cr
              0& 0 & 0 &-1\cr}.\quad$$

Using the commutation relations it can be shown that in four  dimensions we
have the following orthonormal bases for the linear subspaces of ${\cal
S}_4$.
$${\cal A}_{(1)}^-=\{a_1,a_2,a_3\},\quad\quad
{\cal A}^-_{(2)}=\{b_1,b_2,b_3\},$$
$${\cal A}^+_{(1)}=\{I\},$$
$${\cal A}_{(2)}^+=\{c_1,c_2,e_1\},\quad\quad
  {\cal A}_{(3)}^+=\{p_1,q_2,d_2\},\quad\quad
  {\cal A}_{(4)}^+=\{p_2,q_1,d_1\},\quad\quad $$
$${\cal A}_{(5)}^+=\{c_1,p_1,p_2\},\quad\quad
  {\cal A}_{(6)}^+=\{c_2,q_2,q_1\},\quad\quad
  {\cal A}_{(7)}^+=\{e_1,d_2,d_1\},\quad\quad $$

Orthonormal bases for linear subspaces of ${\cal S}_8$ can be constructed using
the sets given above. For example,
the choice $B_s\in \{d_2,p_1,q_2\}$ determines the
possible choices for $B_a$'s, $A_a$'s and $D_a$'s and leads to the CDFN plane.
On the other hand the choice $B_s=I$ leads to the plane obtained by conjugation
given above.

We now show that the basis obtained by choosing $B_s=I$ corresponds to the
representation of ${\cal C}l_7$ using octonionic multiplication.
 Let us describe an octonion by a pair of
quaternions $(a,b)$. Then the octonionic multiplication  rule is
$(a,b) \circ (c,d)=(ac-\bar{d}b,da+b\bar{c})$. If we represent an octonion
$(c,d)$ by a vector in $R^8$, its multiplication by imaginary octonions
correspond to linear transformations on $R^8$. Using the multiplication rule
above, it is easy to see that we have the following correspondences:
$$(i,0)\to\pmatrix{b_1&0\cr0&-b_1\cr},\quad\quad
  (j,0)\to\pmatrix{b_2&0\cr0&-b_2\cr},\quad\quad$$
$$(k,0)\to\pmatrix{b_3&0\cr0&-b_3\cr},\quad\quad
  (0,1)\to\pmatrix{  0&  I\cr  -I&0\cr},\quad\quad$$
$$  (0,i)\to\pmatrix{  0&a_1\cr a_1&0\cr},\quad\quad
  (0,j)\to\pmatrix{  0&a_2\cr a_2&0\cr},\quad\quad
  (0,k)\to\pmatrix{  0&a_3\cr a_3&0\cr}.\quad\quad$$

Finally, we would like to point out that
these constructions can be generalised
to dimensions which are multiples of eight, by replacing unit element with
identity matrices of appropriate size.

In dimensions which are multiples of $16$, one can make use of the
property ${\cal C}l_{k+8}={\cal C}l_k\otimes {\cal C}l_8$ to obtain  a ${\cal
C}l_{k+8}$ representation on $R^{16n}$, using  an already known representation
of ${\cal C}l_k$ on $R^n$.
Hence linear subspaces of ${\cal S}_{16n}$ can be obtained
from the knowledge of
the linear subspaces of ${\cal S}_n$.

\vskip 4mm

\noindent {\bf 5. Conclusion}
\vskip 2mm

In this paper we have characterised  (anti)self-dual
Yang-Mills fields in even dimensional spaces by putting constraints on the
eigenvalues of $F$. The previously known cases of self-dual
Yang-Mills fields in four and eight dimensions are consistent with our
characterisation.
We believe this new approach to self-duality in higher dimensions
deserves further study. It might appear more important to try to understand
the totality of the non-linear space of self-dual 2-forms
as the choice of a linear
subspace of ${\cal S}_{2n}$ is a priori incidental.
Nevertheless, there are
some exceptional linear subspaces, probably the most
important being the one in eight dimensions given by
octonionic multiplication. In this way the close connection
between the self-dual gauge fields in eight dimensions and the
octonionic instantons ${}^{[12],[13],[14]}$ becomes self-evident.

\vskip 2cm
\noindent {\bf Acknowledgement}
\vskip 3mm
We thank Professors A. Trautman and A. D. Popov for  comments.
\newpage
\noindent {\bf References}
\vskip 5mm
\begin{description}
\item{[1]} A.H.Bilge, T.Dereli, \c{S}.Ko\c{c}ak,
 Lett.Math.Phys. {\bf 36}, 301 (1996)
\item {[2]} A. Trautman, Int. J. Theo. Phys. {\bf 16}, 561 (1977)
\item {[3]} D. H. Tchrakian, J. Math. Phys. {\bf 21}, 166 (1980)
%\\ T.N.Sherry, D.H.Tchrakian, Phys.Lett.{\bf B147}(1984)121
%\\ D.H.Tchrakian, Phys.Lett.{\bf B150}(1985)360
%\\ C. Sa\c{c}l{\i}o\u{g}lu, Nucl.Phys.{\bf B277}(1986)487
\item{[4]} T.A.Ivanova, A.D.Popov, Theo. Math. Phys. {\bf 94}, 225 (1993)
\item{[5]}  B.Grossman, T.W.Kephart, J.D.Stasheff,  Commun. Math.
Phys., {\bf 96}, 431 (1984).
(Erratum, ibid {\bf 100}, 311 (1985))

\item {[6]}   E.Corrigan, C.Devchand, D.B.Fairlie and J.Nuyts,
Nuclear Physics {\bf B214}, 452 (1983).
\item{[7]} N.Steenrod, {\bf The Topology of Fibre Bundles} (Princeton U.P. , 1951)
\item{[8]} S.Kobayashi,K.Nomizu, {\bf Foundations of Differential Geometry}
Vol.II ( Interscience , 1969)
\item{[9]} F.W.Warner,{\bf Foundations of Differentiable Manifolds and Lie Groups}\\
(Scott and Foresman, 1971)
\item{[10]} H. Blaine-Lawson,Jr., M.-L.Michelsohn,
{\bf Spin Geometry} (Princeton U.P., 1989)
\item{[11]} E. Witten,  Phys.Lett.{\bf 77B}, 394-400, (1978)
\item {[12]} D.B.Fairlie,J.Nuyts, J. Phys.{\bf A17}(1984)2867
\item {[13]} S.Fubini,H.Nicolai,Phys. Lett.{\bf 155B}(1985)369
\item {[14]} R.D\"{u}ndarer,F.G\"{u}rsey,C.-H.Tze,
 Nucl. Phys.{\bf B266}(1986)440

\end{description}
\end{document}